\documentclass{webofc}
\usepackage[varg]{txfonts}   
\newcommand{\non}{\nonumber\\}
\usepackage[bitstream-charter]{mathdesign}
\usepackage{enumitem}
\usepackage{mathtools}
\begin{document}
\title{Subleading power corrections to heavy quarkonium production in QCD factorization approach}
%
%

\author{\firstname{Kyle} \lastname{Lee}\inst{1,2,3} \and
        \firstname{Jian-Wei} \lastname{Qiu}\inst{4,5} \and
        \firstname{George} \lastname{Sterman}\inst{6} \and
         \firstname{Kazuhiro} \lastname{Watanabe}\inst{7,8}\fnsep\thanks{\email{kazuhiro-watanabe@st.seikei.ac.jp}} 
}

\institute{
Nuclear Science Division, Lawrence Berkeley National Laboratory, Berkeley, CA 94720, USA
\and
Physics Department, University of California, Berkeley, CA 94720, USA
\and
Center for Theoretical Physics, Massachusetts Institute of Technology, Cambridge, MA 02139, USA
\and
Theory Center, Jefferson Lab, Newport News, Virginia 23606, USA
\and
Department of Physics, The College of William \& Mary, Williamsburg, Virginia 23187, USA
\and
C.N.~Yang Institute for Theoretical Physics and Department of Physics and Astronomy, 
Stony Brook University, Stony Brook, NY 11794, USA
\and
SUBATECH UMR 6457 (IMT Atlantique, Universit\'e de Nantes, IN2P3/CNRS), 4 rue Alfred Kastler, 44307 Nantes, France
\and 
Faculty of Science and Technology, Seikei University, Musashino, Tokyo 180-8633, Japan
}

\abstract{
	We report the current understanding of heavy quarkonium production at high transverse momentum ($p_T$) in hadronic collisions in terms of QCD factorization. In this presentation, we highlight the role of subleading power corrections to heavy quarkonium production, which are essential to describe the $p_T$ spectrum of quarkonium at a relatively lower $p_T$. We also introduce prescription to match QCD factorization to fixed-order NRQCD factorization calculations for quarkonium production at low $p_T$.
}
\maketitle
%
\section{Introduction}
\label{intro}

Heavy quarkonium production allows us to study fundamental QCD dynamics about how a quark and antiquark pair is produced and then
fragments into a color singlet hadron, one of the central issues discussed in this conference. 
Thus far, many theoretical efforts have been made to push the accuracy of calculations in NRQCD factorization~\cite{Bodwin:1994jh} toward the NLO level and beyond, and to apply the NRQCD (and pNRQCD) factorization approach to help pin down the quarkonium production mechanism~\cite{Brambilla:2022ayc}. We have argued that the lack of a systematic treatment of significantly enhanced $\ln(p_T^2/m_Q^2)$-type corrections to many NRQCD calculations could affect data fitting of long-distance-matrix-elements (LDMEs) to describe the shape of the $p_T$ spectrum and polarization of heavy quarkonium simultaneously.

We studied the $p_T$ spectrum of heavy quarkonium production employing the QCD factorization approach, which expands the production cross sections in powers of $1/p_T$ first and factorizes both the leading power (LP) and next-to-leading power (NLP) contributions in terms of perturbatively calculable partonic scattering, convoluted with universal parton distribution functions (PDFs) and fragmentation functions (FFs), respectively~\cite{Lee:2021oqr}. The factorized LP cross-section has successfully described LHC data on $J/\psi$ production at high $p_T\gg m_{J/\psi}$~\cite{Lee:2021oqr}. Here we show that the NLP contribution is required to describe $J/\psi$ production data for $p_T  \gtrsim \mathcal{O}(m_{J/\psi})$, whose shape differs from the LP distribution.

In this presentation, we highlight the role of the NLP corrections to hadronic $J/\psi$ production by comparing our calculations with the LHC and Tevatron data to demonstrate that the NLP contribution starts to dominate over the LP one when $p_T\lesssim 4m_{J/\psi}$. We also will provide the justification of an assumption, described below, that makes the QCD evolution equations of the twist-4 double parton FFs more manageable. Finally, we briefly discuss how to match the NLP cross-section to the fixed order calculation in NRQCD factorization.

\section{QCD factorization formalism for inclusive quarkonium production}
\label{sec:qcd-factorization}

Inclusive production of heavy quarkonium $H$ at high $p_T$ ($\gg m_{H}$) in hadronic collisions ($A+B$) can be studied in QCD factorization formalism~\cite{Nayak:2005rt,Kang:2014tta}:
\begin{align}
	E_p\frac{d\sigma_{A+B\to H(p)+X}}{d^3p}
	&\approx E_p\frac{d\sigma_{A+B\to H(p)+X}}{d^3p}\bigg|_{\rm LP} + E_p\frac{d\sigma_{A+B\to H(p)+X}}{d^3p}\bigg|_{\rm NLP},
	\label{eq:resum}\\
	E_p\frac{d\sigma_{A+B\to H(p)+X}}{d^3p}\bigg|_{\rm LP}
	&= \sum\int \frac{dz}{z^2}D_{f\to H}(z,\mu^2)
	E_c\frac{d\hat{\sigma}_{A+B\to f(p_c)+X}}{d^3p_c}\left(p_c=\frac{p}{z},\mu^2\right), \label{eq:LP}
	\\
	E_P\frac{d\sigma_{A+B\to H(p)+X}}{d^3p}\bigg|_{\rm NLP}&=
	\sum_{\kappa}\int \frac{dz}{z^2} D_{[Q\bar{Q}(\kappa)]\to H}(z,\mu^2)
	E_c\frac{d\hat{\sigma}_{A+B\to [Q\bar{Q}(\kappa)](p_c)+X}}{d^3p_c}\left(p_c=\frac{p}{z},\mu^2\right),
	\label{eq:NLP}
\end{align}
where the second and third lines, respectively, represent the factorization formalism at LP and NLP. $d\hat{\sigma}_{A+B\to f(p_c)+X}$ denotes the partonic cross section for producing the fragmenting parton of flavor $f$ and momentum $p_c$ with all collinear sensitivities around $p_c\sim p/z$ absorbed into the twist-2 $f$-to-$H$ FFs, $D_{f\to H}$ with momentum fraction $z$. Similarly, $d\hat{\sigma}_{A+B\to [Q\bar{Q}(\kappa)](p_c)+X}$ denotes the partonic cross section to produce the fragmenting $Q\bar{Q}$ pair of spin-color state $\kappa$ and momentum $p_c = P_Q+P_{\bar{Q}} = P'_Q+P'_{\bar{Q}}$, where $P'_Q$ and $P'_{\bar{Q}}$ are momenta in the conjugated production amplitude. All collinear sensitivities around $p_c$ in the cross-section are absorbed into the twist-4 $Q\bar{Q}(\kappa)$-to-$H$ FFs, $D_{[Q\bar{Q}(\kappa)]\to H}$.

Since physical cross sections should not depend on the factorization scale, the twist-2 and twist-4 FFs satisfy the following coupled evolution equations~\cite{Kang:2014tta},
\begin{align}
	\frac{\partial}{\partial \ln\mu^2} D_{f\to H}(z,\mu^2)
	&= \frac{\alpha_s(\mu)}{2\pi}\sum_{f'}\int_z^1\frac{dz'}{z'}P_{f\to f'}\left(\frac{z}{z'}\right)D_{ f' \to H}(z',\mu^2)
	\non
	&+\frac{\alpha_s^2(\mu)}{\mu^2}\sum_{\kappa}\int_z^1\frac{dz'}{z'} P_{f\to[Q\bar{Q}(\kappa)]}\left(\frac{z}{z'}\right)
	D_{[Q\bar{Q}(\kappa)]\to H}\left(z',\mu^2\right),\label{eq:twist2-evolution}\\
	\frac{\partial}{\partial \ln\mu^2} D_{[Q\bar{Q}(\kappa)]\to H}(z,\mu^2)
	&=\frac{\alpha_s(\mu)}{2\pi}\sum_{n}\int^1_{z}\frac{dz'}{z'} P_{[Q\bar{Q}(n)]\to [Q\bar{Q}(\kappa)]}\left(\frac{z}{z'}\right)\, 
	D_{[Q\bar{Q}(n)]\to H}(z',\mu^2),
	\label{eq:twist4-evolution}
\end{align}
allowing one to use the renormalization group improved QCD factorization at the NLP accuracy.
The first line of Eq.\eqref{eq:twist2-evolution} is the standard DGLAP evolution of the twist-2 FFs
and the second line of Eq.\eqref{eq:twist2-evolution} represents corrections to the evolution from the situation when the fragmenting parton fragments to a $Q\bar{Q}$ pair at the scale $\mu$ and then the pair fragments to the observed quarkonium $H$.  Eq.\eqref{eq:twist4-evolution} represents the linear DGLAP-like evolution equations for twist-4 $Q\bar{Q}(\kappa)$-to-$H$ FFs.  When combined with the NLP partonic cross sections, the solution of Eq.\eqref{eq:twist4-evolution} gives the NLP contribution and is very significant for quarkonium production at low $p_T$. The solution of Eq.\eqref{eq:twist2-evolution}, with the twist-4 $Q\bar{Q}(\kappa)$-to-$H$ FFs from the solution of Eq.\eqref{eq:twist4-evolution}, effectively resum logarithmically enhanced contributions to the cross section when the fragmenting parton fragments into a heavy quark pair at a scale between $[\mu_0 \sim 2m_Q, \mu \sim p_T]$.

The factorization formalism at LP in Eq.\eqref{eq:LP} should dominate the quarkonium production when $p_T\gg m_{H}\approx 2m_Q$ when the LP $\ln(p_T^2/m_Q^2)$-enhanced contributions are significant, and the power corrections in $1/p_T$ are negligible~\cite{Nayak:2005rt}. Although it is formally suppressed by $1/p_T^2$ for producing a 
$Q\bar{Q}$ pair, compared to the LP formalism, contributions from the NLP factorization formalism are enhanced by the fact that a heavy $Q\bar{Q}$ pair is more likely than a single parton to become a quarkonium $H$.  The NLP contribution plays a crucial role in describing quarkonium production cross section at relatively lower $p_T>\mathcal{O}(m_H)$.

In order to obtain Eqs.\eqref{eq:twist2-evolution} and \eqref{eq:twist4-evolution} from corresponding equations and functions in Ref.\cite{Kang:2014tta}, we introduced the modified double parton FFs as $D_{[Q\bar{Q}(\kappa)]\to H}(z,\mu^2)\equiv \int dudv\,\mathcal{D}_{[Q\bar{Q}(\kappa)]\to H}(z,u,v,\mu^2)$ where $u(v)$ denotes a heavy quark's relative longitudinal momentum fraction in the amplitude (conjugated amplitude), and modified the evolution kernels involving the heavy quark pair as 
$P_{[Q\bar{Q}(n)]\to [Q\bar{Q}(\kappa)]}(z)\equiv \int du\int dv\,\mathcal{P}_{[Q\bar{Q}(n)]\to [Q\bar{Q}(\kappa)]}\left(z,u,v,u'=\frac{1}{2},v'=\frac{1}{2}\right)$, $P_{f\to [Q\bar{Q}(\kappa)]}(z)\equiv \mathcal{P}_{f\to [Q\bar{Q}(\kappa)]}\left(z,u'=\frac{1}{2},v'=\frac{1}{2}\right)$, where $u'(v')$ is the momentum fraction of a heavy quark in the amplitude (conjugated amplitude) on the lower virtuality side. The above modification assumes that the same momentum fractions of the heavy quark pair ($u=v=1/2$) are preferred for forming a physical quarkonium.

Next, we explore whether our assumption $u=v=1/2$ can be justified by considering the derivative of double parton FFs with respect to $\mu^2$ (or the slope of their $\mu^2$-evolution):
\begin{align}
	\mathcal{D}'_{\kappa \to n}(z,u,v,\mu^2)&\equiv \frac{2\pi}{\alpha_s}\frac{d\mathcal{D}_{\kappa \to n}(z,u,v,\mu^2)}{d\ln\mu^2}
	\non
	&=\int _z^1\frac{dz'}{z'}\int _0^1du'\int_0^1 dv'\mathcal{P}_{\kappa\to n}\left(\frac{z}{z'},u,v,u',v'\right)\mathcal{D}(z',u',v',\mu^2),
        \label{eq:evo_slope}
\end{align}
where the shape of the double parton FFs in $u,v,z$-space is modeled with a simple test function $\mathcal{D}(z,u,v)$, assuming $\mathcal{D}(z,u,v) \to D_z(z;\alpha,\beta) D_u(u;\gamma) D_v(v;\gamma)$ with
\begin{align}
	D_z(z;\alpha,\beta)=\frac{z^\alpha(1-z)^\beta}{B[1+\alpha,1+\beta]}, \quad 
	D_{u,v}(x;\gamma)=\frac{x^\gamma(1-x)^\gamma}{B[1+\gamma,1+\gamma]},
	\label{eq:FFmodel}
\end{align}
satisfying $\int_0^1 dz\int_0^1 du\int_0^1 dv \mathcal{D}(z,u,v)=1$. The input parameters $\alpha,\beta,\gamma$ in Eq.\eqref{eq:FFmodel} control the shape of the test function. 
Compared to light hadron production, such as $\pi$ meson, heavy quarkonium production is more sensitive to the information of FFs at large $z$, like parton-to-open heavy flavor FFs. We imitate such FFs by combining a large $\alpha$ and a small $\beta$ for $D_z$. We fix $\alpha=30$, $\beta=0.5$ as implemented in Ref.\cite{Lee:2021oqr}. In our analysis, we take $\gamma=10$ to have a relatively narrow peak around $u=v=1/2$ as input.

\begin{figure}[!htbp]
	\begin{center}
		\includegraphics[width=0.495\textwidth]{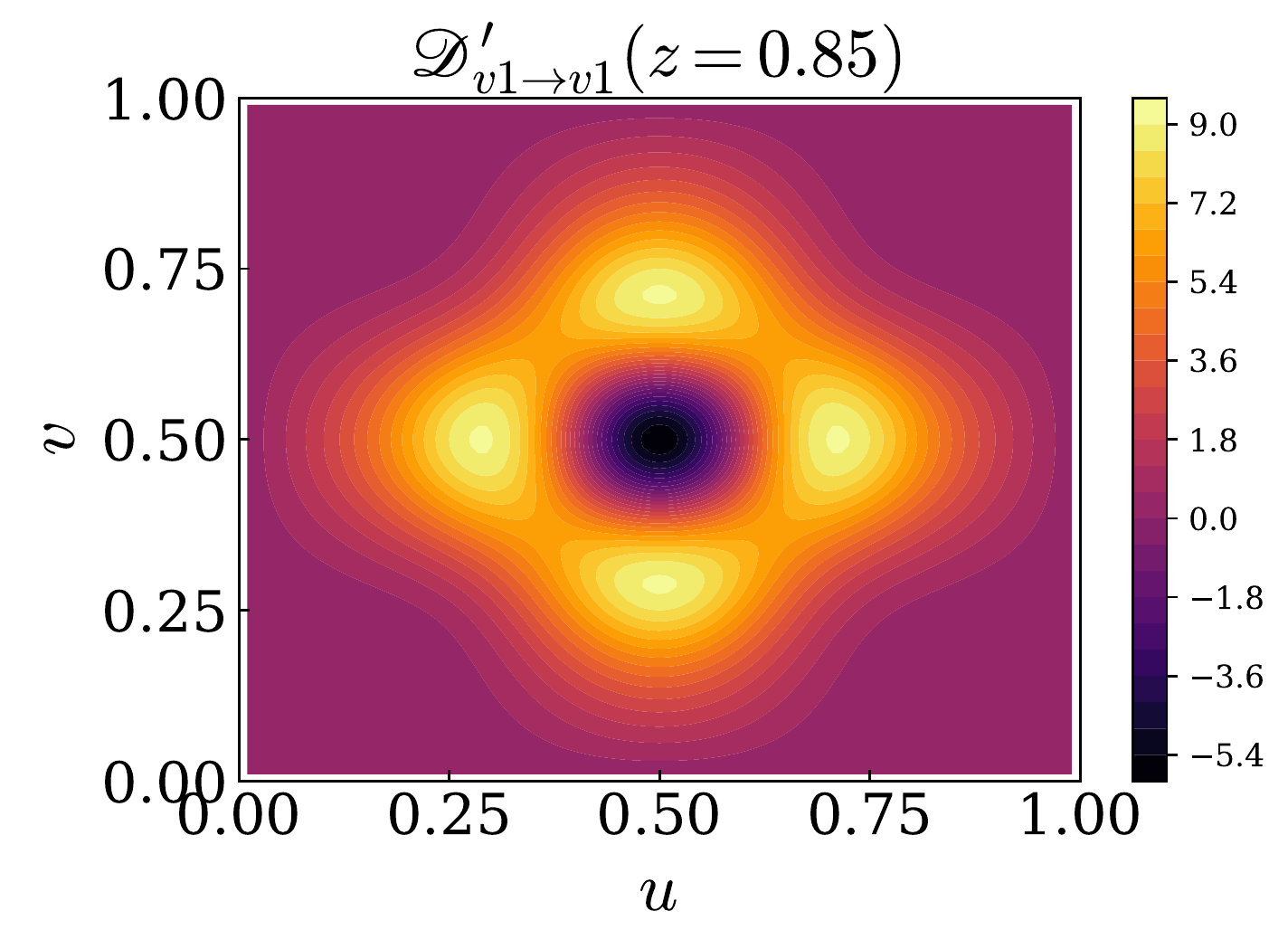}
		\includegraphics[width=0.495\textwidth]{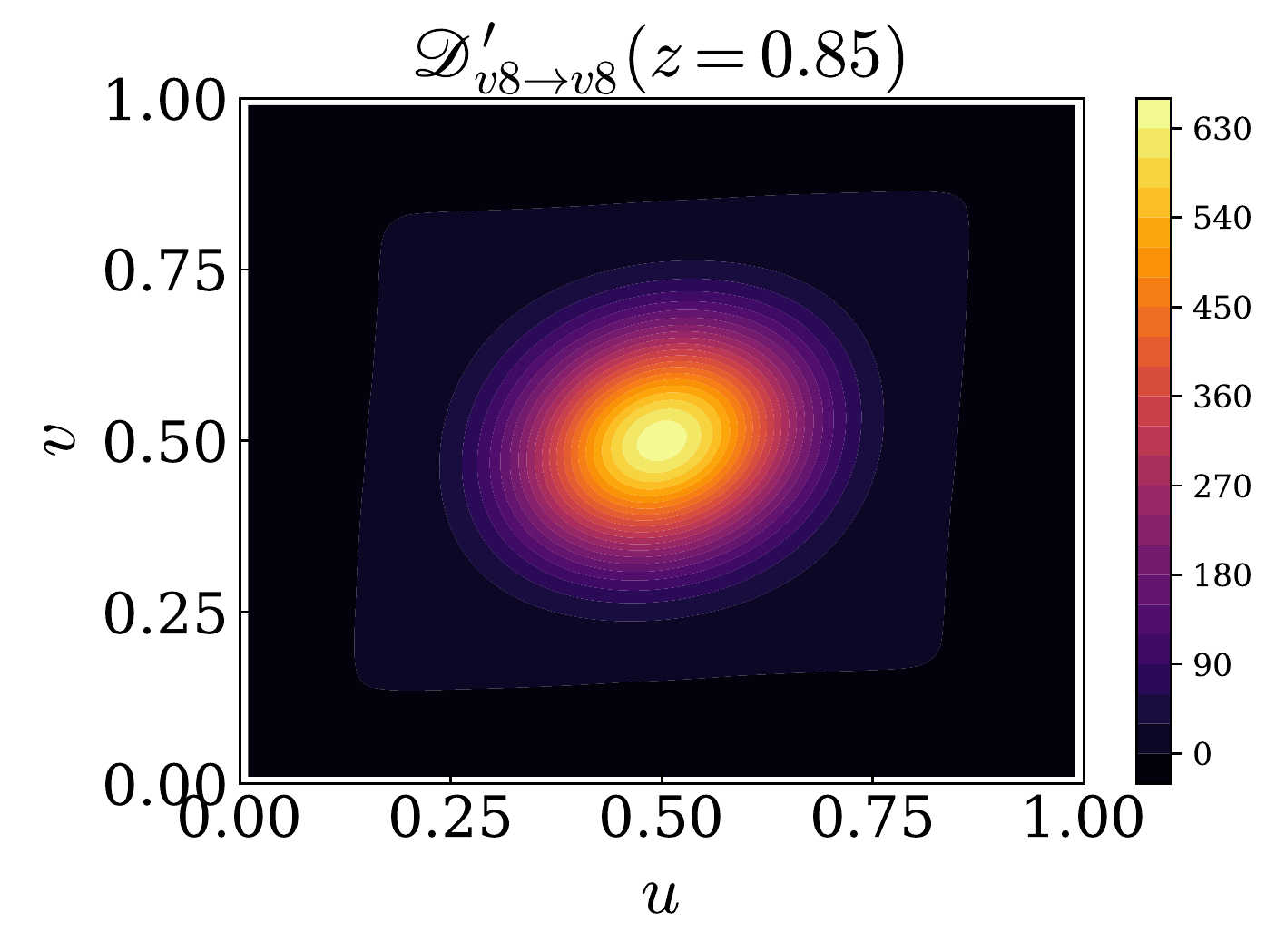}
		\caption{
			Contour plots of $\mathcal{D}'_{v1\to v1}$ (Left) and $\mathcal{D}'_{a8\to a8}$ (Right) for $z=0.85$ in $(u,v)$-space. The sidebar on each panel represents the numerical value of the derivative in arbitrary units; the lighter the sidebar's color, the larger the derivative's value.
		}
		\label{fig:derivative_diagonal}
	\end{center}
	\vspace{-0.1in}
\end{figure}

Figure \ref{fig:derivative_diagonal} shows $\mathcal{D}'_{\kappa \to n}(z,u,v)$ in Eq.\eqref{eq:evo_slope} for the diagonal channels, $\mathcal{D}'_{v1\to v1}$ and $\mathcal{D}'_{v8\to v8}$. For the color-singlet to color-singlet (S-S) diagonal transition, the derivative is negative around $u,v\sim 1/2$ and positive at $u,v\ll 1/2$ and $u,v\gg 1/2$, indicating that the double parton FFs can get broadened in $(u,v)$ space after much evolution. Consequently, our assumption $u'=v'=1/2$ in the evolution kernels is not a good approximation in this case. Nevertheless, since the production rate of quarkonium from a heavy quark pair in the color singlet is much lower than the color octet contribution at high $p_T$, we expect that setting $u'=v'=1/2$ in the evolution kernels would be a good approximation for phenomenological study of the NLP effects.

In contrast, for the color-octet to color-octet (O-O) diagonal channel, the FF can become narrower with a prominent peak around $u=v=1/2$ after much evolution because the derivative at large $z$ is positive over the whole range in $(u,v)$ space, as shown in Fig.\ref{fig:derivative_diagonal}. Therefore, our assumption $u'=v'=1/2$ in the evolution kernels is a good approximation in the large $z$ region. The off-diagonal O-S and S-O channels have the same features as the O-O diagonal channels, so that we can safely approximate $P_Q=P_{\bar{Q}}=P'_Q=P'_{\bar{Q}}=p/(2z)$ in the NLP partonic cross section~\cite{Kang:2014tta}, which has been assumed and used in all perturbative NRQCD calculations.  More details of our analysis will be reported in Ref.\cite{LQSW:2022}.

\begin{figure}[!htbp]
	\centering
	\includegraphics[width=\textwidth]{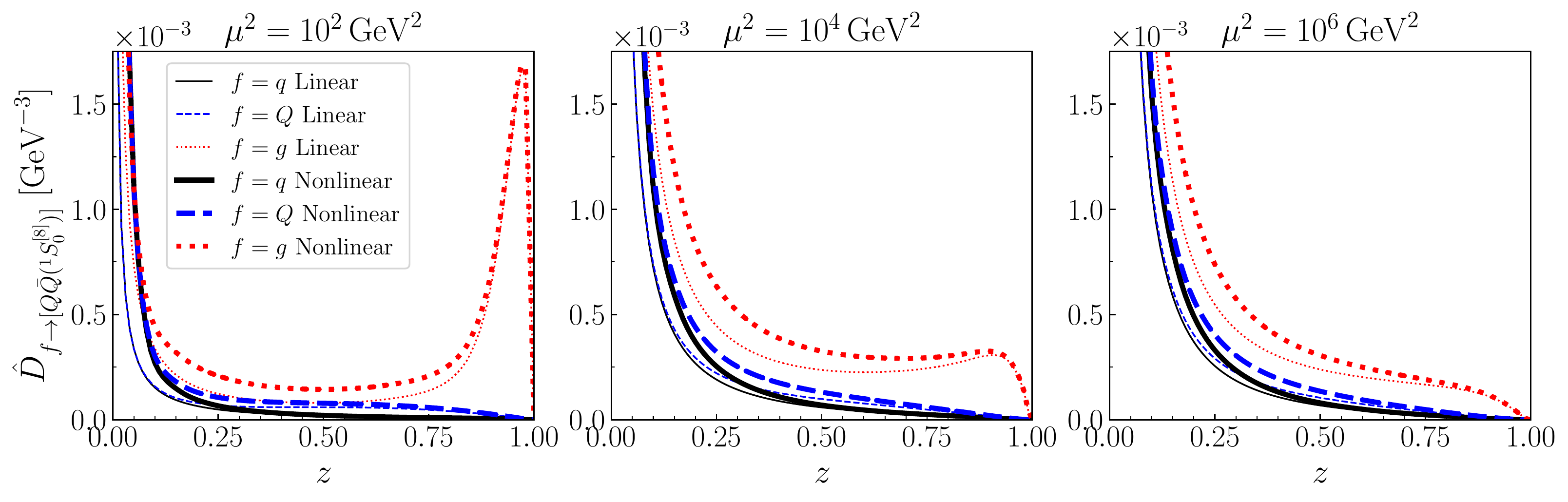}
	\includegraphics[width=\textwidth]{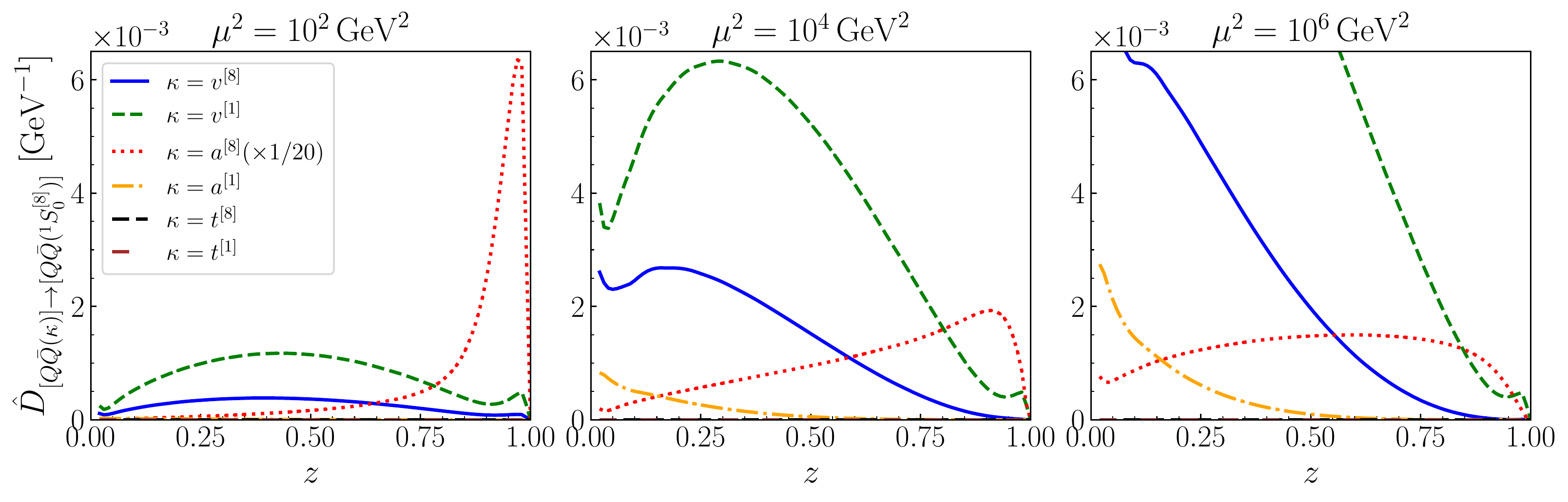}
	\caption{The single and double parton FFs for producing a heavy quark pair in the intermediate state $
	{^1{S}_0^{[8]}}$ as functions of $z$, evolved from $\mu_0^2=(4m_c)^2$ with $m_c=1.5\,{\rm GeV}$ up to given values of $\mu^2$: $\mu^2=10^2\,{\rm GeV}^2$ (left column), $\mu^2=10^4\,{\rm GeV}^2$ (middle column), $\mu^2=10^6\,{\rm GeV}^2$ (right column). $\hat{D}_\kappa=D_\kappa/\langle \mathcal{O}_\kappa\rangle$ with $\langle \mathcal{O}_\kappa\rangle$ the LDME for the production of a quark pair in $\kappa$. The single parton FFs with the linear (thin) and nonlinear (thick) evolution differ in line width in each panel. Black solid curves: light quarks ($u,d,s$), Blue dashed curves: heavy quarks ($c$), Red dotted curves: gluon.}
	\label{fig:FFs_z}
	\vspace{-0.1in}
\end{figure}

\section{Numerical results for hadronic $J/\psi$ production }
\label{sec:result}

We shall show numerical results of hadronic $J/\psi$ production cross section at high $p_T$ in the QCD factorization approach with Eqs.\eqref{eq:LP}-\eqref{eq:twist4-evolution}. We use CT18NLO set for colinear PDFs in the proton and antiproton~\cite{Hou:2019efy}. Significant uncertainty in our approach is from the non-perturbative information of the twist-2 and twist-4 FFs. As studied in Ref.\cite{Lee:2021oqr}, we could approximate these FFs by applying NRQCD factorization to obtain analytic forms of the twist-2 and twist-4 FFs at the input scale, $\mu_0=\mathcal{O}(2m_Q)$, in terms of a finite set of NRQCD LDMEs. The $z$-dependence of the input FFs can be calculated perturbatively in NRQCD in an expansion of $\alpha_s$ and heavy quark velocity $v$ in the pair's rest frame and is available at the first non-trivial order in the powers of $\alpha_s$~\cite{Ma:2013yla,Ma:2014eja}.

Although the non-perturbative input FFs at $\mu_0=\mathcal{O}(2m_Q)\gg \Lambda_{\rm QCD}$ can be approximated by non-perturbative LDMEs with perturbatively calculated coefficients in terms of NRQCD factorization, the perturbative short-distance coefficients inevitably involve very large, even negative, distributions near $z=1$, such as $\delta(1-z)$, $f(z)\ln(1-z)$, $f(z)/[1-z]_+$, and $f(z)[\ln(1-z)/(1-z)]_+$ with $f$ arbitrary finite function of $z$, which are naturally from the perturbative cancellation of IR divergences between real and virtual diagrams. Since heavy quarkonium FFs have peaks at large $z$, these negative terms could have a much more important impact on heavy quarkonium production and drive the perturbatively evaluated cross-section to be negative.  Resummation near the threshold could help us cure this problem of perturbative calculations. At the same time, we must keep in mind that the uncertainty from the expansion of input FFs in terms of $\alpha_s$ and relative velocity $v$, not to be confused with the relative momentum fraction "$v$" above, remains a source of systematic error. We have then taken one particular approach~\cite{Lee:2021oqr}, in which all the perturbatively calculated input FFs are converted into $N_\kappa\, z^\alpha (1-z)^\beta/B[1+\alpha,1+\beta]$, where $\alpha$ and $\beta$ are free parameters, $B$ is the Euler Beta-function, and $N_\kappa$ is equal to the first moment of the related term that takes into account the relative size of different terms from perturbative calculations~\cite{Ma:2013yla,Ma:2014eja}. In our numerical calculations, we have set $\alpha=30$, $\beta=0.5$ for the input FFs at $\mu_0=6\,{\rm GeV}$. We show in Fig.\ref{fig:FFs_z} the $z$-dependence of the single parton and double parton FFs for producing a heavy quark pair of ${^1S_0^{[8]}}$ at different values of $\mu^2$.  As we explained in Ref.\cite{Lee:2021oqr}, the subleading quark pair corrections to the twist-2 evolution equations enhance the twist-2 single parton FFs for a wide range of $z$, as shown in Fig.\ref{fig:FFs_z}, which lead to 10--30\% enhancement to the cross sections even at a sizeable probing scale of $p_T=\mathcal{O}(100\,{\rm GeV})$. The same feature of the nonlinear corrections was found when studying nonlinear corrections to the DGLAP evolution of collinear PDFs~\cite{Mueller:1985wy}.  

\begin{figure}[t]
	\centering
	\includegraphics[width=0.625\textwidth]{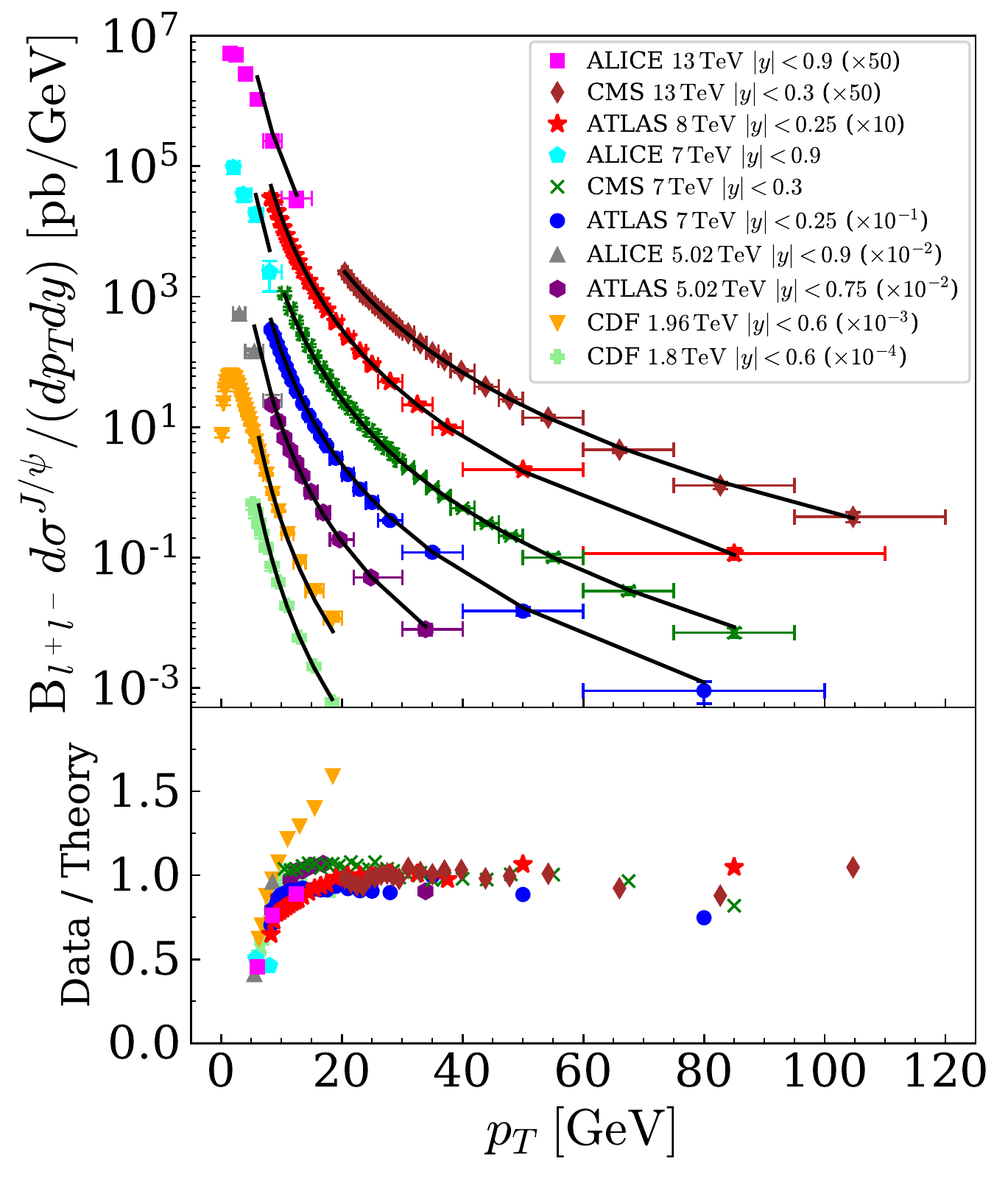}
	\caption{(Upper): Theoretical calculations for the differential cross section of prompt $J/\psi$ production at mid rapidity in hadronic collisions at collider energies. Experimental data are taken from Refs.\cite{CDF:1997ykw,CDF:2004jtw,ALICE:2012vpz,ATLAS:2015zdw,ATLAS:2017prf,CMS:2015lbl,CMS:2017dju,ALICE:2021edd}.
		(Lower): Ratios of mean points of collider data to theoretical predictions in the same $p_T$ bin for prompt $J/\psi$ production, obtained from the upper panel. 
	}
	\label{fig:prompt-Jpsi-colliders}
	\vspace{-0.1in}
\end{figure}

In Ref.\cite{Lee:2021oqr}, we have compared our numerical results of $J/\psi$ production from the QCD factorization formalism with CMS data on prompt $J/\psi$ production at $p_T\ge 60\,{\rm GeV}$ in $\sqrt{s}=7,\,13\,{\rm TeV}$ pp collisions in the rapidity bin $|y|<1.2$~\cite{Khachatryan:2015rra,Sirunyan:2017qdw}, and fixed the overall normalization factor of the theory curve embedded in the input FFs. Our particular setup there 
leads to the production rate of $J/\psi$ being dominated by the ${^1S_0^{[8]}}$ channel, along with some kind of cancelation between other channels.  However, it is important to note that our results are only sensitive to the input FFs from the perturbative QCD factorized single-parton and double-$Q\bar{Q}[\kappa]$ states with $\kappa=v^{[1,8]}, a^{[1,8]}, t^{[1,8]}$, and not sensitive to the details of how each NRQCD channel, such as ${^1S_0^{[8]}}$, ${^3S_1^{[8]}}$, ..., along with the LDMEs, contributes to these input FFs. In this presentation, we extend our analysis to other LHC data and those from Tevatron experiments, for which many precise data sets on hadronic $J/\psi$ production at low $p_T$ down to $p_T=\mathcal{O}(2m_Q)$ are available. 
Without adjusting any normalization and parameters, in Fig.\ref{fig:prompt-Jpsi-colliders}, we compare our calculations to all published data from the LHC and Tevatron on hadronic prompt $J/\psi$ production in the mid rapidity.  The comparison in Fig.\ref{fig:prompt-Jpsi-colliders} clearly demonstrates that the QCD factorization approach with both LP and NLP contributions can describe the data on prompt $J/\psi$ production at collider energies.  With the same choice of $\alpha=30$, $\beta=0.5$ and $K_{\rm NLP}=2$, the K-factor for the NLP contribution that was evaluated at LO in $\alpha_s$ only, the ratio of data over theory calculations is of the order of unity for a wide range of $p_T$.  Clearly, a global fit of the input FFs can further improve the theoretical calculations.

\begin{figure}[t]
	\centering
	\includegraphics[width=0.625\textwidth,clip]{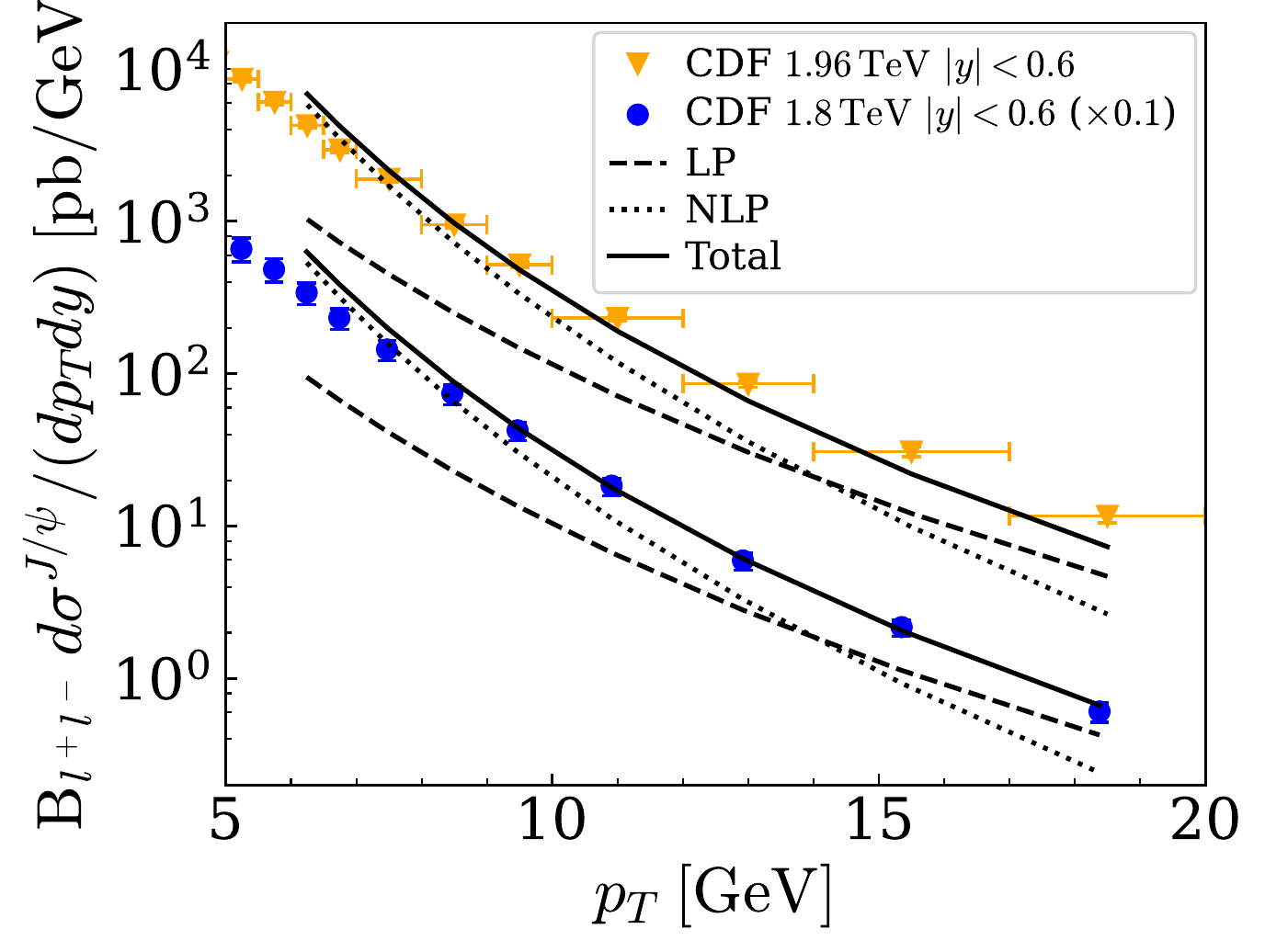}
	\caption{Theoretical calculations for the differential cross section of prompt $J/\psi$ production at mid rapidity in hadronic collisions at Tevatron energies.  Experimental data from CDF are taken from Refs.~\cite{CDF:1997ykw,CDF:2004jtw}.	
	}
	\label{fig:Jpsi-Tevatron}
	\vspace{-0.1in}
\end{figure}

In Fig.\ref{fig:Jpsi-Tevatron}, we focus on the comparison between our calculations with the Tevatron data at relatively low $p_T$ to explore the role of LP and NLP contributions.  The LP curves (dashed lines) include the quark pair corrections to the evolution equations. The NLP contributions (dotted lines) are evaluated with the evolved $Q\bar{Q}$-FFs and the NLP partonic cross sections at the LO along with a $K$-factor, $K_{\rm NLP}=2$, to mimic NLO contributions.  The NLP contribution becomes predominant over the LP contribution around $p_T=15\,{\rm GeV}$ and below at Tevatron energies. 

We expect the LP factorization formula in Eq.\eqref{eq:LP} to work for $J/\psi$ production when $p_T \gtrsim 10 m_{J/\psi}$ where the $\ln(p_T^2/m_c^2)$-type contributions start to dominate. Although the factorized LP contribution provides a good description of the published LHC data at high $p_T$, its extrapolation to lower $p_T$ is below the data at collider energies. The underestimation by the LP contribution emphasizes the importance of the NLP contribution because it is more likely to get the quarkonium from a fragmenting $Q\bar{Q}$-pair than a single parton~\cite{Kang:2014tta}.  Including the NLP contribution provides a better description of data at collider energies even at $p_T\sim 3 m_{J/\psi}$. The LP and NLP contributions distinctly show different $p_T$ spectra; the high $p_T$ data favor the LP curve, while the NLP curve is essential to reproduce the lower $p_T$ data. 

We notice that the NLP contribution starts to overshoot the data points when $p_T\lesssim 2 m_{J/\psi}$, below which the $\ln(p_T^2/(2m_Q)^2)$ is not large, and its contribution cannot dominate the production cross sections.  We need to match our calculations to fixed-order calculations in the NRQCD factorization or pNRQCD factorization~\cite{Brambilla:2022ayc}.  We can implement the matching between the QCD factorization and the NRQCD factorization by defining~\cite{Kang:2014pya}
\begin{align}
	d \sigma_{A+B\to H+X}&=
	d\sigma_{A+B\to  H+X}^\textrm{QCD-Res} 
	+ d\sigma_{A+B\to H+X}^\textrm{NRQCD-Fixed} 
	-d\sigma_{A+B\to H+X}^\textrm{QCD-Asym}
	\non
	&\Rightarrow
	\begin{dcases}
		d\sigma_{A+B\to H+X}^\textrm{QCD-Res}                   
		&\textrm{when}~\quad p_T\gg m_{H};~d\sigma^\textrm{NRQCD-Fixed}\approx d\sigma^\textrm{QCD-Asym}\\
		d\sigma_{A+B\to H+X}^\textrm{NRQCD-Fixed}   
		&\textrm{when}~\quad p_T\to m_{H};~d\sigma^\textrm{QCD-Res}\approx d\sigma^\textrm{QCD-Asym} 
	\end{dcases},
	\label{eq:new-factorization-formula}
\end{align}
where $d\sigma_{A+B\to  H+X}^\textrm{QCD-Res}$ in Eq.\eqref{eq:new-factorization-formula} represents the QCD factorization formalism in Eq.\eqref{eq:resum} including the resummation of $\ln(p_T^2/m_Q^2)$-type contributions, $d\sigma_{A+B\to H+X}^\textrm{NRQCD-Fixed}$ represents the fixed order NRQCD (or pNRQCD) factorization formalism that does not include the $\ln(p_T^2/m_Q^2)$-type resummation, and $d\sigma_{A+B\to H+X}^\textrm{QCD-Asym}$ is equal to the fixed-order expansion of $d\sigma_{A+B\to  H+X}^\textrm{QCD-Res}$. 
The $d\sigma_{A+B\to H+X}^\textrm{QCD-Asy}$ term is necessary for avoiding double-counting and for a smooth transition between the high and low $p_T$ regimes.   This matching has to happen around $p_T\lesssim 3 m_{J/\psi}$ for $J/\psi$ production at Tevatron and the LHC energies.  

\section{Summary}
\label{sec:summary}

We discussed the calculations of $p_T$ distribution of prompt $J/\psi$ production in hadronic collisions at Tevatron, and the LHC energies in the renormalization group improved QCD factorization approach. 
While we focused on the LP contribution to $J/\psi$ production at high $p_T\ge 60$~GeV where
the $\ln(p_T^2/m_Q^2)$-type contribution dominates in Ref.\cite{Lee:2021oqr}, we highlighted in this presentation the importance of the NLP contribution in describing the data at relatively lower $p_T$.  The NLP partonic cross section with the twist-4 double parton FFs was calculated with the assumption of $u=v=1/2$, which was justified for the leading contributions with O-O and O-S color transitions.   We also presented a numerical comparison between our calculations and data from the LHC and Tevatron, showing the consistency for $p_T\gtrsim 3m_{J/\psi}$.  We also introduced a matching formalism for extending theory calculations to $p_T\lesssim 3m_{J/\psi}$, and leave the details of this matching for future work~\cite{LQSW:2022}.

\bigskip
\begin{acknowledgement}
We thank Hee Sok Chung and Xiang-Peng Wang for their valuable discussions.
K.W. would also like to thank Jefferson Lab for the computational resources essential to perform this project. This work is supported by Jefferson Science Associates, LLC, under U.S. DOE Contract No.\, DE-AC05-06OR23177. The work of K.L. and G.S. was supported in part by the National Science Foundation, award 1915093.
\end{acknowledgement}


\bibliography{references}

\end{document}